\documentclass[aps,pre,twocolumn,groupedaddress,showpacs]{revtex4-1}

\usepackage{amsmath}
\usepackage{enumerate}
\usepackage{array}
\usepackage{graphicx}
\usepackage{epstopdf}
\usepackage{amsfonts}
\usepackage[table,xcdraw]{xcolor}
\usepackage{xcolor}

\usepackage{multirow}
\usepackage{hhline}
\usepackage[caption=false]{subfig}

\newcommand{\tp}{\cellcolor[rgb]{0.72,0.74,0.76}}
\newcommand{\tf}{\cellcolor[rgb]{0.82,0.84,0.86}}
\newcommand{\tk}{\cellcolor[rgb]{0.92,0.94,0.96}}

\definecolor{tp}{rgb}{0.72,0.74,0.76}

\begin{document}
\arrayrulecolor{white}
\arrayrulewidth=2pt
%\preprint{}

\title{Memory and burstiness in dynamic networks}


\author{Ewan R. Colman}
\email[]{E.Colman@Reading.ac.uk}
\author{Danica Vukadinovi\'c Greetham}
%\thanks{}
\affiliation{Centre for the Mathematics of Human Behaviour, Department of Mathematics and Statistics, University of Reading, RG6 6AX, UK}


\date{\today}

\begin{abstract}
%We introduce a class of complex network models which evolve through the addition of edges between nodes selected randomly according to their intrinsic fitness, and the deletion of edges according to their age. We add to this a memory effect where the attractiveness of a node is increased by the number of edges it is currently attached to, and observe that this creates burst-like activity in the attachment events of each individual node which is characterised by a power-law distribution of inter-event times. The fitness of each node depends on the probability distribution from which it is drawn; we find exact solutions for the expectation of the degree distribution for a variety of possible fitness distributions, and for both cases where the memory effect either is, or is not present. This work can potentially lead to methods to uncover hidden fitness distributions from fast changing, temporal network data such as online social communications and fMRI scans.

A discrete-time random process is described which can generate bursty sequences of events. A Bernoulli process, where the probability of an event occurring at time $t$ is given by a fixed probability $x$, is modified to include a memory effect where the event probability is increased proportionally to the number of events which occurred within a given amount of time preceding $t$. For small values of $x$ the inter-event time distribution follows a power-law with exponent $-2-x$. We consider a dynamic network where each node forms, and breaks connections according to this process. The value of $x$ for each node depends on the fitness distribution, $\rho(x)$, from which it is drawn; we find exact solutions for the expectation of the degree distribution for a variety of possible fitness distributions, and for both cases where the memory effect either is, or is not present. This work can potentially lead to methods to uncover hidden fitness distributions from fast changing, temporal network data such as online social communications and fMRI scans.

\end{abstract}

%\keywords{Burstiness}

\maketitle

The mathematics of interactive complex systems has a vital role to play in the interpretation of large-scale social and biological data. Technology which facilitates the collection of vast amounts of information is increasingly becoming available for both academic and commercial purposes; however, in the absence of a detailed understanding of the underlying processes, there will always be a risk of deriving the wrong conclusion from the facts. Complexity science provides numerous models of social, biological, physical and economic systems which combine large numbers of individual components to reproduce the types of behaviour observed on the systemic level. The components in such systems are usually uninteresting in isolation, but when allowed to interact with each other they produce complex non-trivial patterns which in some cases agree very well with empirical results. This poses a challenge for data scientists: given information only about the system as a whole, with all its complex and interactive dynamics, how can one conclude anything about the individual components? To begin answering that question we need to understand, in mathematical terms, the form and extent of the biases that complexity creates.

The purpose of the present work is to provide an understanding of how one very simple mechanism, a memory effect (brought about by interaction), will bias the statistical properties of a complex system such as the distribution of communication activity in a social network, or the distribution of brain activity of different cortical regions in a fMRI scan. We consider a hypothetical system of individual agents (nodes) and the instantaneous pairwise interactions which happen between them (edges). By aggregating all of the interactions that occur within some given time window, a network is formed whose structure can be analysed for a deeper understanding of the system. In general, the length of this time window determines the density of the network; as an increasing amount of data is aggregated, a picture of the system emerges which shows not only whether or not two nodes are connected, but also includes the strength of their relationship through the frequency of their interactions.

Throughout this paper we will be comparing two possible forms of stochastic process: Markovian and non-Markovian. In the non-Markovian case the rate of activity of the individual agents in the system is proportional to the number of events which the agent can `remember'; these are events which happened at earlier times and are stored in a memory of a given fixed size. We will refer to this as the ``memory effect''. When a large number of interacting agents are considered, the memory of an individual is recorded in the structure of the network of interactions. Specifically, the number of interactions a node can `remember' is effectively the same as its degree, this way the mechanism for creating links in the non-Markovian network model is a form of linear preferential attachment \cite{Albert}. Likewise, the process of `forgetting' is an edge deletion mechanism \cite{PhysRevE.74.036121}. 

\section{Related work}
The motivation for this work is the growing evidence of memory dependent, burst-like activity in complex interactive systems. Recently this has been most prominent in the study of online communication patterns \cite{gaito2012bursty}. Evidence for burstiness is found in the distribution of inter-event times between actions; in a Poisson process, for example, which is Markovian i.e. memoryless, the inter-event times follows an exponential distribution; periodic events, such as a heartbeat, have inter-event times which generally stay close to the mean; and lastly, in systems which are generally said to be ``bursty'', the inter-event times follow a power-law distribution. A formal definition has been proposed to quantify these behaviours in \cite{goh2008burstiness}. A slightly different approach in \cite{karsai2012universal} identifies a ``burst'' as a sequence of events where each event follows the previous one within a given time interval. This definition naturally leads to the consideration of two possible types of event: those which happen spontaneously, and those which occur as reactions to previous events (e.g. the dynamics of human conversations). 

The current explanations for why a sequence of events might have a power-law inter-event time distribution rely on a memory effect; in other words the probability of an event occurring at a given time is dependent on events which occurred at previous times. Models have been proposed based on queuing theory where incoming messages are replied to according to some prioritisation strategy \cite{barabasi2005origin, gonccalves2011modeling, PhysRevE.83.056101}. By adjusting a parameter which controls the randomness of the strategy, these models have been shown to create power-law distributed inter-event times with exponents that agree with a number of real-world data-sets. 

Many of of the systems in which burstiness has been observed cannot be considered to have the internal mechanisms of a queuing model, these include studies of the human brain \cite{beggs2003neuronal}, animal movement patterns \cite{reynolds2011origin} and consumer behaviour \cite{deschatres2005dynamics}. Bursts of activity closely resemble cascading events such as avalanches and mass extinctions and therefore might possibly be examples of self-organised criticality (SOC) \cite{bak1997nature}, where the focus is on the \emph{emergence} of scale-free distributions based on very few assumptions about the system. In fact, the Bak-Sneppen model \cite{bak1993punctuated}, one of the fundamental examples of SOC, is known to have a power-law distribution of inter-event times \cite{peng2014punctuated}. Much of the analytical progress made in the study of bursts has come from related models \cite{gabrielli2007invasion,vazquez2006modeling,vajna2013modelling,1742-5468-2013-02-P02006,Vazquez2007747}. 
\newline

The present work examines how the bursty behaviour of individual interacting agents affects the large scale macroscopic view of the system. We consider activity on a dynamic network which is closely related to several previously studied models: Preferential attachment \cite{price76, Albert} is a non-Markovian method by which many networks grow. In this process, nodes are added to the network sequentially in discrete time-steps and edges are created between the new node and old nodes selected randomly but with probability proportional to their degree. The rate of growth in connectivity of a node at any given time therefore depends on its entire history. Conversely in ``fitness'' networks the connectivity of a node accords only to an attractiveness value drawn from some probability distribution \cite{PhysRevLett.89.258702}. Such models are versatile in their applications as they can incorporate various topological network features such as clustering, and have also allowed complex network topologies to be incorporated into SOC models \cite{garlaschelli2007self} (we note here that a significant proportion of the fitness network literature concerns correlations between the fitness of connected nodes, while the present work concerns only uncorrelated networks). A simple way to combine fitness and preferential attachment has been achieved by defining the attractiveness of a node to be either the sum or product (or a combination of both) of its degree and its intrinsic fitness \cite{dorogovtsev2000structure,bianconi2001competition,ergun2002growing}.

The networks mentioned so far are static in the sense that once a link is created between nodes it remains in that location forever. In many situations this is not the case and we here use the term ``dynamic'' to refer to networks whose edges can be removed as new ones are created \cite{PhysRevE.74.036121,ben2007addition,evans2007exact,colman2014local}. The model introduced in \cite{xie2008scale} combines the preferential selection of nodes with an added fitness parameter (the same for every node) on a dynamic network where edges are removed so that the total number of edges remains constant. The authors focus on the problem of finding the degree distribution; what they do not mention is that the degree of each individual node in this model fluctuates with a memory-driven bursty process with power-law distributed inter-event times between each attachment event. The present work provides a mathematical description of this behaviour. Additionally, by incorporating heterogeneous fitness distributions into the previous model, we will describe a class of complex networks which exhibit a rich variety of structural and time-dependent properties.
\newline

The model presented in this paper is a versatile and applicable dynamic network with varying node fitness. There is currently research activity in related areas that is of much interest: time varying networks, in particular, have some similarities with dynamic networks. Informally speaking, these are multi-layered networks where each layer corresponds to a distinct time interval, they differ from dynamic networks because at the end of each time interval the entire network (rather than just a single edge) is removed and replaced \cite{perra2012activity,hoppe2013mutual}. In its most basic form the ``action potential'' (the propensity to act at any given time) of each node does not change with time.  In \cite{karsai2014time} memory effects are considered within the time varying formulation. The authors observe in social communication data a universal rule for the probability that an individual will continue an old correspondence rather than start a new one. Adding this constraint to the original time-varying concept gives accurate results regarding the number of contacts and the weight of correspondence with each contact.

In \cite{moinet2014burstiness} the waiting time distribution between actions takes an arbitrary form, thus the action potential of each node may vary. When the waiting time takes a power-law form they find that the exponent of the degree distribution depends on the exponent of the waiting time distribution. Similarly, in the model introduced in \cite{vestergaard2014memory}, the rates at which new links are formed and broken, and the rate at which old links become active, depend on the probability distribution of inter-event times. The authors choose to examine the power-law inter-event time distribution, commenting that this is akin to a preferential attachment mechanism. Unlike the present work however, the power-law is an assumption and not an emergent property.

The aforementioned studies do not contradict the work presented here, these papers are complimentary. Together they reinforce the movement to unify bursty dynamics and network structure. 
\newline

The remainder of this paper is structured as follows: We describe a process for generating a sequence of events which, under certain circumstances, produces a power-law inter-event time distribution. In the section which follows, we introduce a model of an evolving network where edges are removed and replaced at each time-step. Within this section two possible attachment kernels are described, the first is entirely fitness based, the second has an additional preferential attachment mechanism which can be interpreted as an increased propensity to act caused by previous interactions. We show that in the latter case, the activity of the nodes is described by the random process of Section \ref{sequence}. Results are presented and we present figures which show the degree distributions in some special cases of the model. In Section \ref{discussion} we highlight the advances achieved by this research. We then discuss briefly its possible applications and elements which require further study. In Appendix \ref{iet_solution} the solution for the inter-event time distribution is shown. In Appendix \ref{derivation} we show how the network is described mathematically, and derive results regarding the degree distribution for a general fitness distribution. In Appendix \ref{examples} we look at some special cases of the fitness distribution. 

\section{Generating event sequences with power-law distributed inter-event times}
\label{sequence}
Before discussing the network topology of a population of interacting agents, we first examine a process which describes the memory dependent behaviour of an individual node. We describe a discrete-time stochastic process which generates an infinite sequence of binary random variables $X_{1},X_{2},\ldots$. At time $t$ we may have $X_{t}=1$, which we consider to be an `event', or $X_{t}=0$, which we consider to be a moment of inactivity. The system has a memory capacity of size $M$, meaning that there are $M$ locations, $m_{1}(t),m_{2}(t),...,m_{M}(t)$, where an event may be stored, i.e. $m_{n}(t)\in 0,1$ for $n\in1,2,...M$. We define $k_{t}=\sum_{n=1}^{M}m_{n}(t)$ and let the event probability kernel $f$ be any function such that $0<f(k_{t})<1$ for $k_{t}\in 0,...,M$. We consider two possible ways, random and age-based, in which events can be deleted from the memory.
 \begin{figure}[b]
 \includegraphics[width=0.5\textwidth]{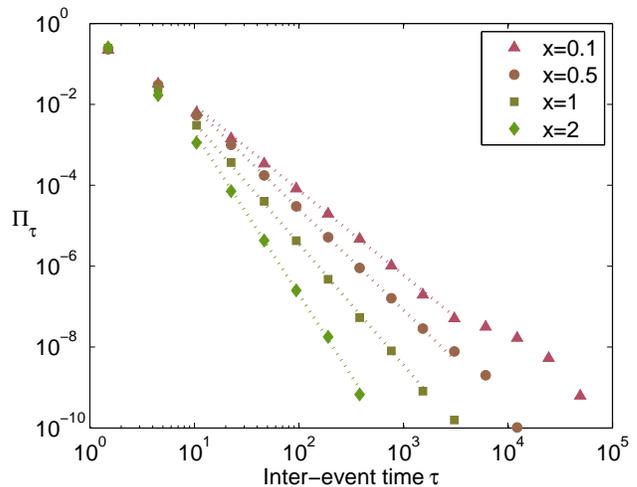}
  \caption{\label{iet_fig}The inter-event time distribution for a sequence of events generated by the process described in Section \ref{sequence_rand}. The simulation lasted for $10^{8}$ iterations with $M=10^{3}$ and $\epsilon=1$. The markers show the log-binned frequencies (normalised to give the proportion of inter-event times of length $\tau$). The dotted lines show the corresponding slope predicted by Eq.\eqref{pi_tau}.}
\end{figure}

\subsection{Randomised memory}
\label{sequence_rand}
At time $t$,
\begin{enumerate}
\item With probability $f(k_{t})$, $X_{t}=1$. With probability $1-f(k_{t})$, $X_{t}=0$.
\newline

\item Integer $n'$ is selected uniformly at random from $1,2,...,M$ and $m_{n'}(t+1)=X_{t}$. For all other $n\neq n'$, $m_{n}(t+1)=m_{n}(t)$.
\end{enumerate}
Since there is always a non-zero probability that $X_{t}=1$ (and similarly that $X_{t}=0$) the process will continue indefinitely without ever reaching an absorbing state. For example, if on the contrary $f(k_{t})=k_{t}/M$, and we start from an initial state where $k_{0}\neq 0$, then we will eventually end up in one of two states: either $k_{t}=M$ or $k_{t}=0$. Analysis of this particular case is important to evolutionary biology \cite{moran1962statistical}. We find that by eliminating the possibility of absorption the statistical properties of the sequence can be calculated in the $t\rightarrow \infty$ limit. 
\subsection{Age-based memory}
\label{sequence_age}
The randomised memory process is approximately equivalent to the following alternative description: In each iteration we perform step $1$ as before. We then set $m_{M}(t+1)=X_{t}$ and $m_{n}(t+1)=m_{n+1}(t)$ for $1\geq n \geq M-1$. This way $i$ will `remember' all the events which happened in the previous $M$ iterations. For example,
\begin{equation}
\ldots0000010\hspace{-2px}\underbrace{1010010}_{M}X_{t}. \nonumber
\end{equation}
Assuming that the value of $m_{1}$ is not correlated with the value of $k_{t}$, i.e. the probability of removing a $1$ is well approximated by $k_{t}/M$, then the solutions given in Appendix \ref{iet_solution} are applicable in both cases.
\newline

We are interested in $\Pi_{\tau}$, the probability that the inter-event time of a randomly selected pair of consecutive events will be exactly $\tau$. Our analysis focuses on the linear probability kernel 
\begin{equation}
\label{kernel_0}
f(k_{t})=\frac{k_{t}+x}{M+x+\epsilon}
\end{equation}
where $x$ and $\epsilon$ are real positive numbers. As $x$ increases, the contribution from the memory factor $k$ becomes less important and the system approaches a Bernoulli process. At the other extreme, when $x$ is small relative to $M$, the inter-event time distribution asymptotically follows a power-law. In Appendix \ref{iet_solution} we derive an approximate solution to $\Pi_{\tau}$ showing that the exponent of the power-law is independent of the parameters $M$ and $\epsilon$, but is dependent on the choice of $x$ in the following way
\begin{equation}
\label{pi_tau}
\Pi_{\tau}\sim \tau^{-(2+x)}.
\end{equation}
Numerical results are presented in Fig.(\ref{iet_fig}) for a range of values of $x$. The deviation away from a power-law that is present in the very large values of $\tau$ can be attributed to the fact that once the waiting time reaches such high values, it becomes overwhelmingly likely that $k=0$, meaning that memory effects are null. 
\newline

Section \ref{dyn_mem} concerns a network of agents who create edges with other nodes dynamically according to a preferential attachment process, and destroy edges either randomly or according to their age. After introducing the network model we will show that its parameters can be equated with the parameters of the stochastic process described in this section.  

\section{Dynamic network model}
\begin{figure}[hb]
 \includegraphics[width=0.4\textwidth]{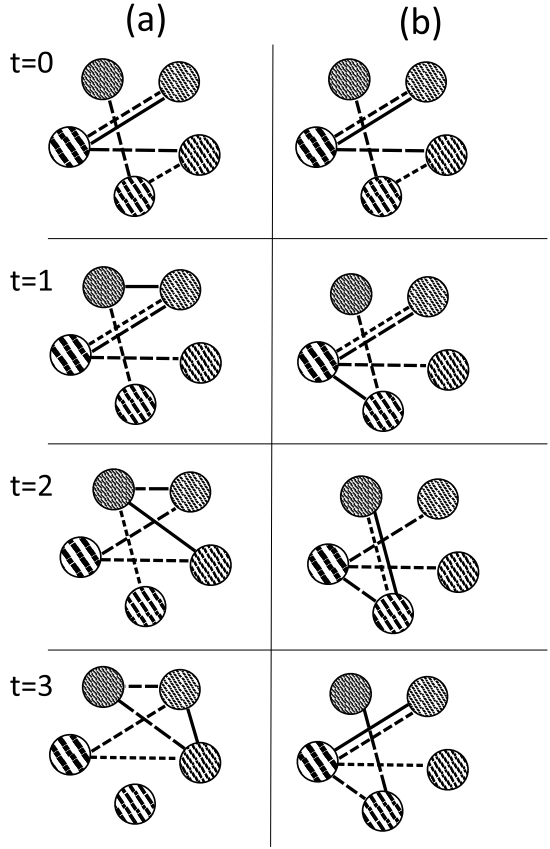}
  \caption{\label{fig:figure1}Three iterations of the network model starting from a random initial configuration. The number of stripes inside each node corresponds to the fitness, the node with the least stripes has fitness $x=0.5$, the others have $x=1$, $x=1.5$, $x=2$, $x=2.5$ and $x=3$. The number of dashes in each edge corresponds to its age with the oldest having the most dashes. In each iteration the oldest edge is removed and a new edge is added between nodes selected either with probability proportional to their fitness (shown in (a)) or with probability proportional to the sum of their fitness and their degree (shown in (b)). Note that in (a) the fittest node is also the most active. In (b) this is not the case.}
\end{figure}

We consider a network formed of $N$ nodes and $E$ edges. Initially the edges are placed between pairs of randomly selected nodes. For each node, a positive continuous random variable $x\in \mathbb{R}$ is selected according to a probability density function $\rho(x)$ which has mean $\langle x \rangle$. Following the related literature we shall refer to this value as the ``fitness'' of the node, denoted $x_{i}$ for the node $i$. The degree $k_{i}$ is the total number edges adjacent to $i$ (note that this not the same as the number of neighbours of $i$ since multiple edges can exist between any pair of nodes). The dynamics of the system are described as follows: in each iteration, a node $i$ is randomly selected with probability given by its attachment kernel $\Pi(i)$, a second node is selected in the same way and an edge is created between them. In the same iteration the oldest edge is removed (thus $E$, $N$ and the mean degree $\langle k \rangle=2E/N$ remain constant throughout). Alternatively we could choose to remove a randomly selected edge instead of the oldest, these two possibilities correspond to the randomised and age-based forms of the processes described in \ref{sequence_rand} and \ref{sequence_age} respectively; the results presented here are applicable to both. Under these rules, the probability that an edge will be created between two nodes $i$ and $j$ is proportional to the product of their fitness $\Pi(i)\Pi(j)$. This is just one of many ways to combine the fitness of two nodes; a wealth of literature exists examining the other possibilities and generalisations (see for example \cite{PhysRevLett.89.258702,smolyarenko2013network}). The process is illustrated in Fig.(\ref{fig:figure1}) for both of the attachment kernels considered here.

In most real-life situations the fitness of a node represents some hidden (or latent) quantity, whereas its degree represents something tangible that appears in empirical data-sets. In general, then, an important problem to address is in inferring the fitness of the node when given only its degree and other properties describing the structure of the network. In a stochastic system the closest we can get to achieving this is finding the \emph{probability} that a node has fitness $x$ given some information about the network structure. When the available information is the degree of each node, Bayes rule gives the appropriate expression for this quantity: 
\begin{equation}
P(x|k)=\frac{\rho(x)P(k|x)}{p_{k}}
\end{equation}
where $p_{k}$ is the probability of randomly selecting a node which has degree $k$, and $P(k|x)$ is the same probability but this time conditioned on $x$. Thus there is an incentive to extract these quantities; as well as being interesting in their own right; they are integral to uncovering the hidden variables. The analysis in this section focuses mainly on deriving the degree distribution and the conditional degree distribution for a range of fitness functions. We consider the two following possible attachment kernels.
\subsection{Dynamic model without memory}
The probability of attaching one end of an edge to a node $i$ of fitness $x_{i}$ is
\begin{equation}
\label{kernel1}
\Pi(i)=\frac{x_{i}}{\sum_{j}x_{j}}=\frac{x_{i}}{N\langle x \rangle}
\end{equation}
Under this condition the $x_{i}$ can be considered the rate of activity of $i$ and one might naively assume that the relationship between $x_{i}$ and the degree of $i$, $k_{i}$, is approximately linear (specifically $k_{i}\approx x_{i}\times\langle k \rangle/\langle x \rangle$ since this would give the correct result for the total degree of the network). In general, this is not the case; Figures (\ref{expo_forgetful}) and (\ref{power_forgetful}) show that the degree distributions and fitness distributions of networks created by this process after a large number of time-steps contain fundamental differences. If $\rho(x)=\lambda e^{-\lambda x}$ then the degree distribution depends only on the mean degree of the network and not at all on the parameter $\lambda$. In this case there are therefore infinitely many possible fitness distributions which produce the same degree distribution. If $\rho(x)$ follows a power-law with exponent $\gamma$ then $p_{k}$ will have a power-law tail with the same exponent $\gamma$, small values of $k$, however, become increasingly uncommon as we look at denser networks.  

 \begin{figure}[t]
 \includegraphics[width=0.5\textwidth]{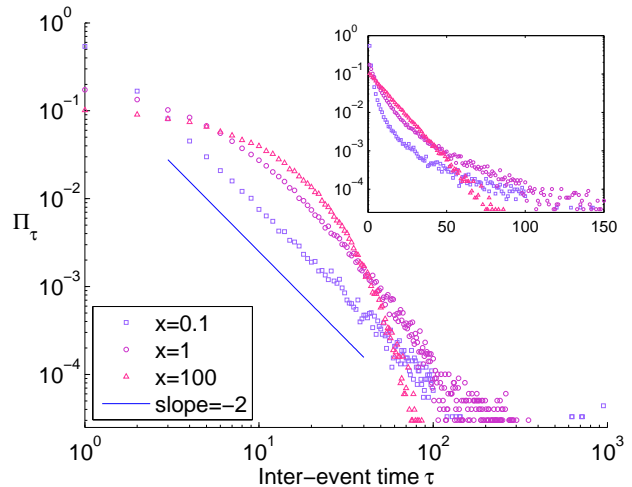}
  \caption{\label{fig:figure2}(Colour online) The inter-event time distribution for attachment events of a single node $i$ with fitness equal to the mean of the fitness distribution ($x_{i}=\langle x\rangle$) on Log-Log axes (main), and Log-Linear axes (inset). The plotted results consider a dense network where $N=10$ and $E=100$. The attachment events of $i$ are described by the process introduced in Section \ref{sequence} with $M$=$E$ and $\epsilon=(N/2-1)\langle x \rangle$. When $x_{i}$, and hence $N\langle x\rangle$, are very large, the event probability is dominated by the contribution from $x_{i}$ and is therefore weakly dependent on the memory of $i$. In this case the inter-event times are distributed 
exponentially (as expected in a Bernoulli process). As $\langle x\rangle\rightarrow 0$ the contribution from the memory of previous events becomes dominant and the distribution approaches a power-law with exponent $-2$.}
\end{figure}

\begin{figure*}[t!]
        \subfloat[Exponential without memory.]{\label{expo_forgetful}
                \includegraphics[width=0.45\textwidth]{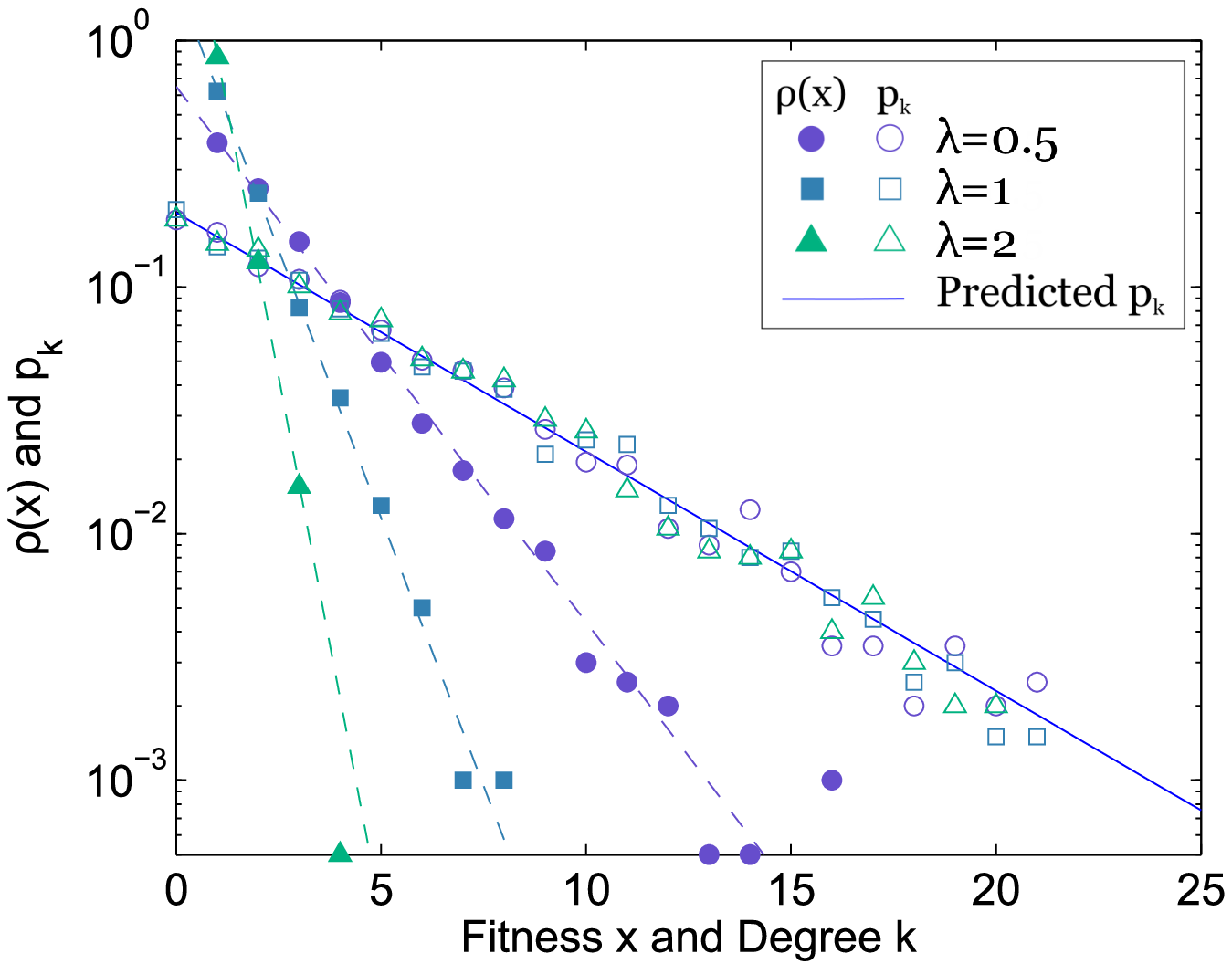}}
                \qquad
        \subfloat[Exponential with memory.]{\label{expo_memory}
                \includegraphics[width=0.45\textwidth]{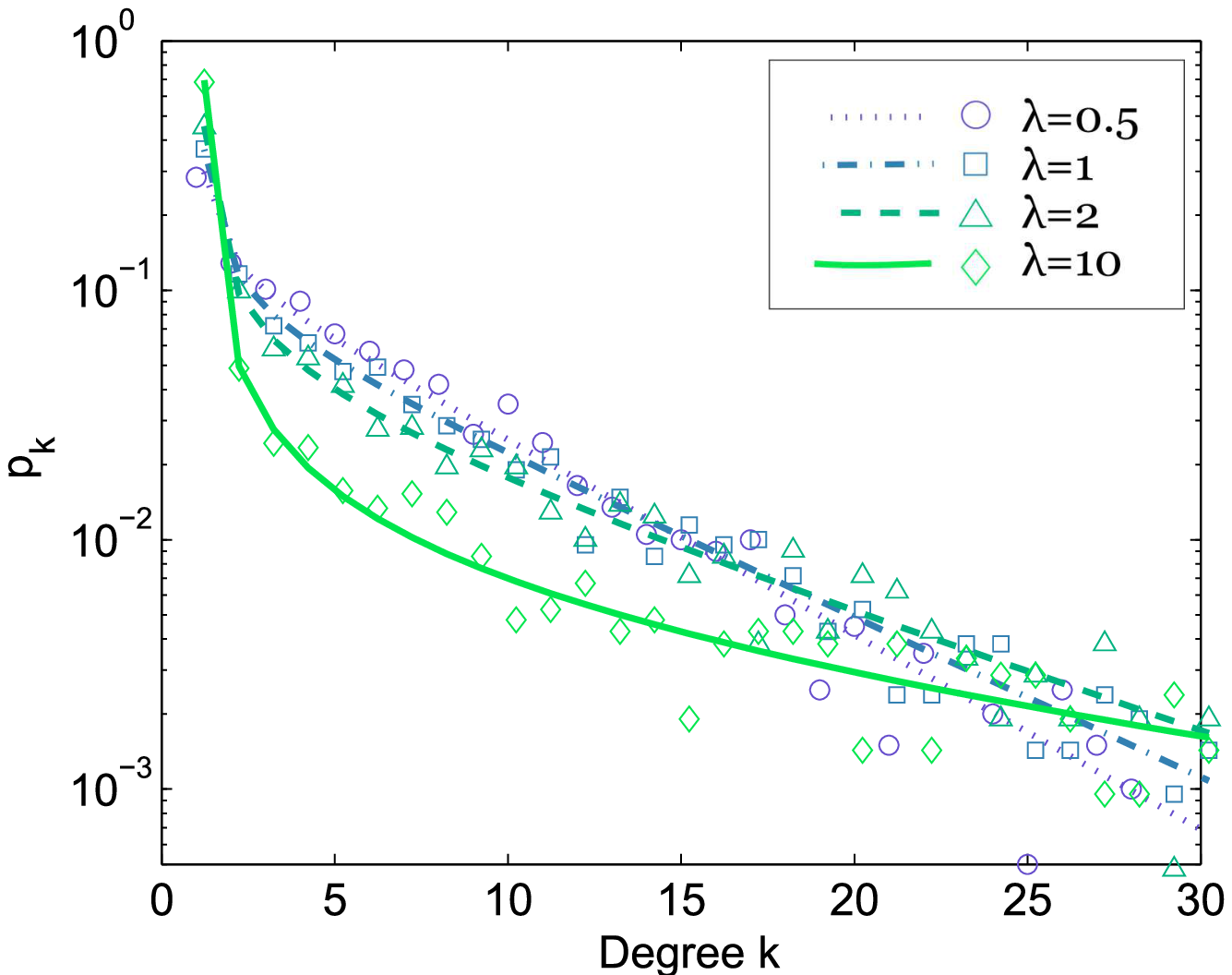}}
				        \qquad
				\subfloat[Power-law without memory.]{\label{power_forgetful}
                \includegraphics[width=0.45\textwidth]{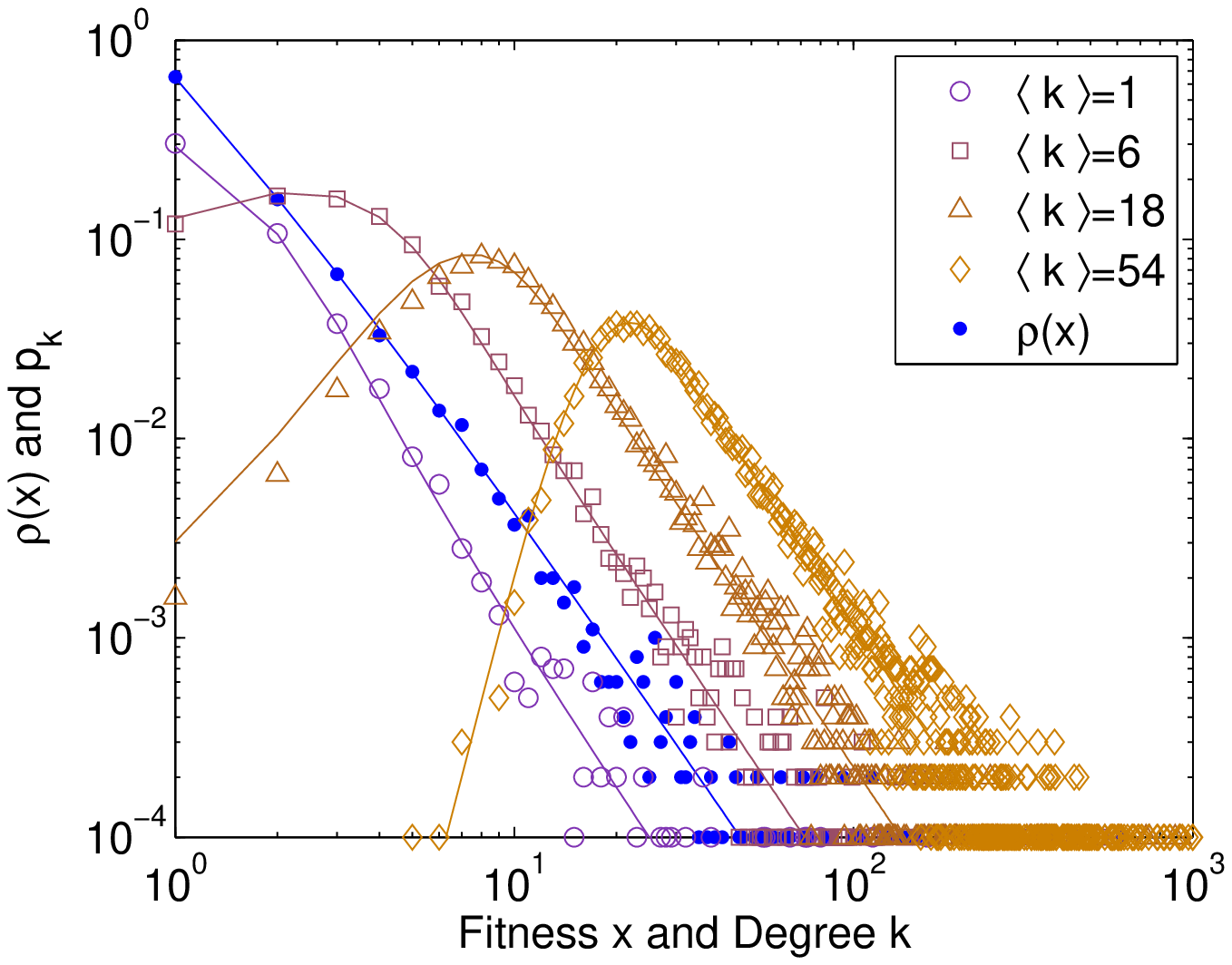}}				        
                \qquad
        \subfloat[Power-law with memory.]{\label{power_memory}
                \includegraphics[width=0.45\textwidth]{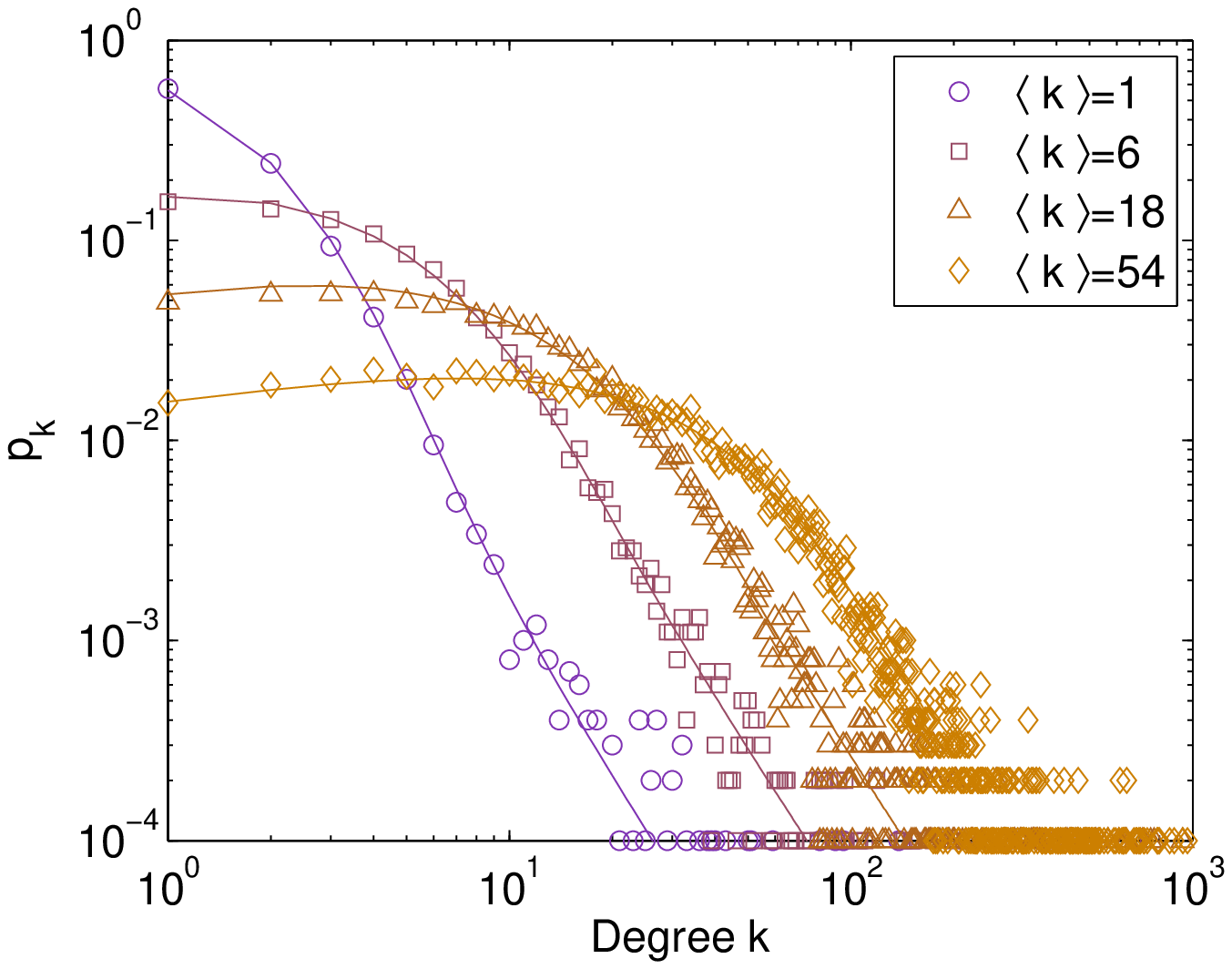}}               
 \caption{\label{fig:figure0}(Colour online) The degree distributions for both types of attachment kernel and two different forms of $\rho(x)$. In each plot the markers show the results of a single numerical simulation of a network of $2\times 10^{3}$ nodes, the smooth lines show the corresponding analytical results.  In (\ref{expo_forgetful}) $E=4 \times 10^{3}$ and the fitness distribution is $\rho(x)=\lambda e^{-\lambda x}$, which is special case of the gamma distribution [Eq.(\ref{gamma_dist}) with $\alpha=1$ and $\beta=1/\lambda$]. For a range of values of $\lambda$, an exponential fitness distribution is plotted with filled markers and the degree distribution is plotted with unfilled markers of the same shape. We see that in this particular case the parameter $\lambda$ does not effect the result. This is not the case when memory effects are introduced, shown in (\ref{expo_memory}); as $\lambda$ increases the degree distribution approaches a power-law [see Eq(\ref{gamma_result})]. In (\ref{power_forgetful}) the power-law fitness distribution [given by Eq.(\ref{power_dist}) with $x_{\text{min}}=1$ and $\gamma=2.5$] is plotted next to the degree distributions for a range of values of $E$ (giving different densities). For large values of $k$ the degree distribution has the same power-law exponent as the fitness distribution, even when memory effects are introduced as we see in (\ref{power_memory}). The effect of including memory is however seen in the small values of $k$.}
\end{figure*}

\subsection{Dynamic model with memory}
\label{dyn_mem}
The probability of attaching one end of an edge to a node $i$ of fitness $x_{i}$ is
\begin{equation}
\label{memory_kernel}
\Pi(i)=\frac{k_{i}+x_{i}}{\sum_{j}(k_{j}+x_{j})}=\frac{k_{i}+x_{i}}{N(\langle k \rangle + \langle x \rangle)}
\end{equation}
where $k_{i}$ is the degree of $i$. Memory in this system is recorded by the edges, as the current degree influences the creation of future edges. Because the edges in this system are dynamic, in that the oldest one is removed with each iteration while new ones are added, each node effectively has a memory which extends backward in time to the age of the oldest edge. Let us now consider the relationship between this attachment kernel and the process described in Section \ref{sequence}. If we consider the attachment of the end of an edge to the node $i$ to be an `event', then the probability of an event occuring at time $t$ is given by Eq.\eqref{memory_kernel} multiplied by $2$ (corresponding to the two ends of the edge). Additionally, the event will be deleted from memory after precisely $E$ iterations, so the number of edges corresponds to the length of the memory i.e. $M=E$. Therefore, when $\epsilon$ is chosen such that $x_{i}+\epsilon=N\langle x \rangle/2$, Eq.\eqref{kernel_0} and $2$ times Eq.\eqref{memory_kernel} become identical and we conclude that the results of Section \ref{sequence} apply to the sequence of attachment events for individual nodes in this model. It is possible then, by choosing a fitness distribution which ensures that $\langle x\rangle<<\langle k\rangle$, to create burst like patterns of behaviour in the activity of the node $i$. As we deviate away from these conditions the activity of the nodes is better described by a Bernoulli process, giving exponentially distributed inter-event times, seen in Fig.(\ref{fig:figure2}). 

Results for the degree distribution are plotted in Figs.(\ref{expo_memory}) and (\ref{power_memory}). We find that in the case where the fitness follows a gamma distribution, $p_{k}$ approaches a power-law with exponent $-1$ as the mean fitness $\langle x \rangle\rightarrow 0$. In this limit, the heterogeneity in the degree distribution can be explained entirely by the fluctuations seen in individual nodes, given by Eq.\eqref{pk_approx}; the fact that one node might have a greater fitness than another node becomes irrelevant. This however is not the case when the fitness distribution follows a power-law. If $\rho(x)$ follows a power-law with exponent $\gamma$ then $p_{k}$ will have a power-law tail with the same exponent $\gamma$. The effect of introducing memory is seen mostly in the smallest values of $k$ which, in contrast to the memoryless case, remain relatively frequent, even in dense networks. We conclude then, that while both gamma and power-law distributions have tails which extend to infinity, as $\langle x \rangle\rightarrow 0$ the effect of memory dominates over the fitness distribution in the first case, and the fitness distribution dominates over the effect of memory in the second. 

Given the degree distribution from a system whose behavior meets the description of the model, our analysis suggests a method to infer the hidden fitness distribution numerically by assuming it takes the form of a step function, reducing the problem to an optimisation problem given in Appendix \ref{examples}.

\section{Discussion}
\label{discussion}
This analysis has potential to be useful in many applications. Suppose we have a system where data arrives in the form of a list of interactions between a finite number of agents. This model provides a framework for interpreting such data. A sample of say, $n$ interactions, can be thought of as a network with $n$ edges, all of which are placed according to some hidden fitness variable which the present model makes no assumptions about. It should also be noted that the assumption in our model that interactions are pairwise can easily be generalised so that any given number of nodes may be active at each time-step (i.e. a hypergraph). We have shown the impact of the edge density (which can be interpreted as the size of the sample) on the degree distribution and that the effect of bursty, memory driven, behaviour is seen mostly in the nodes which have low intrinsic fitness. We also suggested a method to recover the hidden fitness distribution from the data. We note that the variability with edge density is very similar to the problem of time varying networks discussed in \cite{perra2012activity} although in this work the authors focus on the issue of not counting multiple edges more than once (something the present analysis ignores) and are not concerned with the aggregate network after a long time when it reaches a high density. The results of this paper have shown that the effect at high densities is profound and can be significantly altered by the addition of memory.

The motivation for this work was the potential applicability to two specific areas of data analysis: online social interactions (e.g. Twitter) and the data received from fMRI scans, in particular when the cortical regions of the brain are considered as nodes and activity may transmit from one region to another (see for example \cite{grindrod2013primary,petri2014homological}). Both systems are known to exhibit bursty activity. In the case of human communication this is brought about by the reciprocation of messages. Empirical studies \cite{vazquez2006modeling} have found the power-law exponent in the inter-event time distribution to be between $-1$ and $-2$. This behaviour can be recreated by our model but it requires negative values of $x$ and for $f(0)$ in Eq.\eqref{kernel_0} to be defined separately. Further work would therefore be required to make this analysis directly applicable. Less is known about why burstiness occurs in the human brain but it is likely because of some kind of feedback mechanism \cite{bak2001adaptive}. Recovering a fitness distribution using the dynamic model with memory in either of these situations would effectively amount to filtering out the effects of these internal feedback mechanisms and exposing the external influences on the system.

Our final remark is a mention of the burst pattern result observed in the activity of nodes (Fig.(\ref{fig:figure2})). In many studies of burstiness (such as \cite{moinet2014burstiness} and \cite{vestergaard2014memory}) the power-law inter-event time distribution is included as an \textit{a priori} assumption in the description of the model. We have shown that this pattern can emerge from a simpler, lower-level process, suggesting that there could be a universal reason why such patterns are observed so frequently in complex systems. The relationship between this result and the well studied SOC models needs to be established in order to move towards an analytical understanding of both phenomena, hopefully broadening this model to a wide range of universality classes, and potentially extending its applicability. 
\begin{acknowledgments}
This work is funded by the RCUK Digital Economy programme via EPSRC Grant
EP/G065802/1 `The Horizon Hub'.
\end{acknowledgments}

\appendix
\section{Solution for the inter-event time distribution}
\label{iet_solution}
We first find $p_{k}$, the probability that $k_{t}=k$ for a randomly selected $t\in \mathbb{N}$. For the general event probability kernel $f$ we find a recursion relation relating $p_{k}$ to $p_{k-1}$. We then continue by examining only the special case where
\begin{equation}
\label{kernel}
f(k)=\frac{k+x}{M+x+\epsilon}
\end{equation}
for constants $x$ and $\epsilon$ and find the exact solution for $p_{k}$. From this result we approximate the probability that the time between two events is exactly $\tau$ iterations of the model. 
\subsection{Memory size distribution}
\begin{table}[h]
\centering
\begin{tabular}{p{12px}ccc}
                                            &\multicolumn{1}{r}{}                 & \multicolumn{2}{ c }{\tp Added} \\ 
                                                          \hline
                                            &\multicolumn{1}{c}{\tp}            &  \multicolumn{1}{l}{\tf $1$}&\multicolumn{1}{l}{\tf $0$}\\
                                            &\multicolumn{1}{c}{\tp Probability}&  \tf  $f(k_{t})$                 & \tf  $1-f(k_{t})$              \\ 
                                            &\multicolumn{1}{c}{\tp }           &  \tf                         & \tf                        \\
                              \hline
\tp                                         &\multicolumn{1}{l}{\tf $1$}        &  \tk                         & \tk                        \\
\tp                                         &\multicolumn{1}{c}{\tf $k_{t}/M$}      &  \tk  $k_{t+1}=k_{t}$        & \tk  $k_{t+1}=k_{t}-1$     \\ 
\tp                                         &\multicolumn{1}{c}{\tf}            &  \tk                         & \tk                        \\
%\arrayrulecolor{tp} \hhline{-~~~} \arrayrulecolor{white} 
\hline
\tp                                         &\multicolumn{1}{l}{\tf $0$}        &  \tk                         & \tk                        \\
\tp                                         &\multicolumn{1}{c}{\tf $1-k_{t}/M$}    &  \tk  $k_{t+1}=k_{t}+1$      & \tk  $k_{t+1}=k_{t}$       \\ 
\tp\multirow{-6}{*}{\rotatebox{90}{\colorbox{tp}{\hspace{7px} Removed \hspace{7px}}}}&\multicolumn{1}{c}{\tf}            &  \tk                         & \tk                        \\
\end{tabular}
\caption{At each iteration $X_{t}\in0,1$ is added to the memory while at the same time a randomly selected entry will be removed. We show the probabilities of these events and how each possible combination changes $k_{t}$.}
\label{transitions}
\end{table}

Table \ref{transitions} shows the possible events which can happen regarding the addition and deletion of $1$s in the memory. All possible transitions of $k_{t}$ are brought together in the following master equation which describes the evolution of $p_{k}(t)$:
\begin{equation}
\begin{split}
p_{k}(t)=&\left[1-\frac{k-1}{M}\right]f(k)p_{k-1}(t-1)\\
&+\left[1-\frac{k}{M}-f(k)+2\frac{k}{M}f(k)\right]p_{k}(t-1)\\
&+\left[1-f(k+1)\right]\left(\frac{k+1}{M}\right)p_{k+1}(t-1)
\end{split}
\end{equation}
As $t\rightarrow \infty$ the distribution will converge towards a time-invariant distribution, $p_{k}$, described by
\begin{equation}
\label{invariant}
\begin{split}
\left[1-\frac{k-1}{M}\right]f(k)p_{k-1}-\left[\frac{k}{M}+f(k)-2\frac{k}{M}f(k)\right]p_{k}&\\
+\left[1-f(k+1)\right]\left(\frac{k+1}{M}\right)p_{k+1}&=0.
\end{split}
\end{equation}
This \textit{second-order} recurrence relation reduces to a \textit{first-order} recurrence relation with the introduction of 
\begin{equation}
H(k)=\frac{k}{M}\left[1-f(k)\right]p_{k} 
\end{equation}
and
\begin{equation}
 F(k)=f(k)\left[1-\frac{k}{M}\right]p_{k};
\end{equation}
using the condition that $p_{-1}=0$ in Eq.(\ref{invariant}) we see that $F(0)=H(1)$ and also that Eq.(\ref{invariant}) becomes
\begin{equation}
F(k)-F(k-1)=H(k+1)-H(k).
\end{equation}
Clearly then $F(k-1)=H(k)$, and so $p_{k}$ obeys the first-order recurrence equation:
\begin{equation}
\label{rec1}
p_{k}=\left(\frac{M-1-k}{k}\right)\left[\frac{f(k-1)}{1-f(k)}\right]p_{k-1}.
\end{equation}
Writing $p_{1}$ in terms of $p_{0}$, then $p_{2}$ in terms of $p_{1}$ and so on, we can express Eq.(\ref{rec1}) as
\begin{equation}
p_{k}=p_{0}\prod_{i=1}^{k}\left(\frac{M-1-i}{i}\right)\left[\frac{f(i-1)}{1-f(i)}\right].
\end{equation}
We choose at this point to investigate only the linear case with $f(k)$ given by Eq.(\ref{kernel}). In this instance the translation property of the Gamma function ($x\Gamma(x)=\Gamma(x+1)$) can be used and we arrive at
\begin{equation}
\label{pk_exact}
p_{k}=\frac{\Gamma(M-1)\Gamma(M-k+\epsilon)\Gamma(k+x)}{\Gamma(M+\epsilon)\Gamma(M-k-1)\Gamma(x)k!}p_{0}
\end{equation}
giving the probability distribution for the number of $1$s in the memory at any given time. 
\subsection{Inter-event time distribution}
Here we derive $\Pi_{\tau}$, the probability that a randomly selected interval has size $\tau$. Suppose we select a random $X_{t}$ from the sequence. For $X_{t}$ to be $0$ and belong to an interval of length $\tau$ it must be preceded by a string composed of a $1$ followed by $\tau'$ $0$s, and it must be the first $0$ in a sequence of $\tau-\tau'$ $0$s followed by another $1$. The variable $\tau'$ can be any integer from $1$ to $\tau$ and we need to sum the probabilities of each possibility to arrive at the probability that $X_{t}$ is a $0$ at \emph{any} location within an interval of size $\tau$. Expressed symbolically, the previous sentence is equivalent to
\begin{equation}
\label{interval1}
\tau \Pi_{\tau}(t)=\sum_{\tau'=1}^{\tau}f(k_{t-\tau'})f(k_{t+\tau-\tau'+1})\prod_{i=t-\tau'+1}^{t+\tau-\tau'}1-f(k_{i})
\end{equation} 
where $\Pi_{\tau}(t)$ is the probability that the interval containing $X_{t}$ has length $\tau$. The multiplication by $\tau$ on the left hand side comes from the fact that there are $\tau$ choices of $X_{t}$ which belong to this interval. We make the following approximations and coarsening of the model:
\begin{enumerate}
\item We assume that $M$ is large and also consider only the values of $k$ large enough for Stirling's approximation to be a valid to approximate the Gamma functions in Eq.(\ref{pk_exact}). We further limit our attention to those values of $k$ for which $M>>k$ and get
\begin{equation}
\label{pk_approx}
p_{k}\approx\left[1-\frac{k}{M}\right]^{1+\epsilon}\frac{p_{0}}{\Gamma(x)}k^{x-1}\approx\frac{p_{0}}{\Gamma(x)}k^{x-1}
\end{equation} 
\item We choose $M>>\delta,\epsilon$  which means $f(k_{t})\approx k_{t}/M$. More importantly, if we say that $P(f(k_{t})=\phi)$ is the probability that $f(k_{t})=\phi$ for a randomly selected $t\in \mathbb{N}$ then from Eq.(\ref{pk_approx}) we have 
\begin{equation}
P(f(k_{t})=\phi)\approx \frac{p_{0}}{\Gamma(x)}[M\phi]^{x-1}.
\end{equation}
\item Over short time periods, changes to $k_{t}$ will be small. In other words, locally the system behaves as a Bernoulli process with success probability given by $\phi$. This allows Eq.(\ref{interval1}) to be approximated by
\begin{equation}
\Pi_{\tau}(t)=f(k_{t})^{2}[1-f(k_{t})]^{\tau}.
\end{equation}
When $t\in \mathbb{N}$ is selected randomly this is equivalent to
\begin{equation}
\label{cond}
P(\tau|f(k_{t})=\phi)=\phi^{2}(1-\phi)^{\tau}.
\end{equation}
\item We approximate $\phi$ by a continuous variable.
\end{enumerate}
The time-independent solution to the inter-event time distribution is found by solving
\begin{equation}
\begin{split}
\Pi_{\tau}&=\int_{0}^{1}P(f(k_{t})=\phi)P(\tau|f(k_{t})=\phi)d\phi\\
&\approx\frac{p_{0}M^{x-1}}{\Gamma(x)}\int_{0}^{1}\phi^{x+1}(1-\phi)^{\tau}d\phi.
\end{split}
\end{equation}
Thus we find that the inter-event time distribution is given by a Beta function $\Pi_{\tau}\sim B(\tau+1,x+2)$which for large values of $\tau$ obeys
\begin{equation}
\label{result}
\Pi_{\tau}\sim \tau^{-(2+x)}.
\end{equation}

\section{Solution for the network degree distribution}
\label{derivation}
\subsection{Dynamic model without memory}
For each positive integer $k$ we want to know the number of nodes $n_{k}$ that have degree $k$ as a function of the fitness distribution $\rho(x)$, as well as the parameters $N$ and $E$. This quantity is the expectation of the degree distribution; the mean of the ensemble of networks generated in this way. Letting $t$ be the number of iterations and $n_{k}(x,t)$ be the expectation of the number of nodes of degree $k$ with fitness $x$ at time $t$, we can write down the rate of change
\begin{equation}
\label{rate1}
\begin{split}
\frac{\partial n_{k}(x,t)}{\partial t}=&\frac{2x}{N\langle x \rangle}[n_{k-1}(x,t)-n_{k}(x,t)]\\
&+\frac{1}{E}[(k+1)n_{k+1}(x,t)-kn_{k}(x,t)].
\end{split}
\end{equation}
The first two terms on the right hand side account for the creation and destruction (respectively) of nodes of degree $k$ which occurs when an edge is attached to a node of degree $k-1$ (creation) or to a node of degree $k$ (destruction). The last two terms on the right hand side account for the creation and destruction of nodes of degree $k$ which occurs when the oldest edge is removed from a node of degree $k+1$ (creation) or removed from a node of degree $k$ (destruction). We have assumed here that the ages of edges adjacent to a node are not correlated, thus the process of removing the oldest edge is approximately the same as removing a randomly selected one. After a large number of iterations the system will be in equilibrium, $n_{k}(x,t)=n_{k}(x)$, and the left hand side will be equal to zero. Using a similar method to that found in \cite{xie2008scale} we solve Eq.(\ref{rate1}) by introducing
\begin{equation}
H(k,x)=\frac{2x}{N\langle x \rangle}n_{k}(x) \text{ and } G(k,x)=\frac{1}{E}kn_{k}(x).
\end{equation}
Eq.(\ref{rate1}) now becomes
\begin{equation}
\label{GH}
G(k+1)-G(k)=H(k)-H(k-1).
\end{equation}
By summing Eqs.(\ref{GH}) over all $k\geq 1$ we find that $G(0,x)=H(1,x)$ and consequently $G(k,x)=H(k-1,x)$ for all $k\geq 1$, solving this leads to
\begin{equation}
\label{zero}
n_{k}(x)=\left(\frac{2Ex}{N\langle x \rangle}\right)^{k}\frac{1}{k!}n_{0}(x).
\end{equation}
To find $n_{0}$ we consider $N(x)$, the expected number of nodes of fitness $x$, 
\begin{equation}
\label{zero2}
N(x)=\sum_{k=0}^{\infty}(lx)^{k}\frac{1}{k!}n_{0}(x)=n_{0}(x)e^{lx}
\end{equation}
where $l=2E/N\langle x \rangle$. The conditional probability $P(k|x)=n_{k}(x)/N(x)$ is found by combining Eq.(\ref{zero}) and Eq.(\ref{zero2}) to get
\begin{equation}
\label{conditional1}
P(k|x)=\frac{1}{k!}(lx)^{k}e^{-lx}.
\end{equation}
Thus isolating only the nodes which have fitness exactly equal to $x$ we find a Poission degree distribution. Interestingly, this implies that if one was to take a sample of nodes which all have a similar fitness value, one would see a network which looks very similar to an Erd\H{o}s-R\'enyi random graph. [Eq.(\ref{conditional1}) can also be found by more direct means. It can be expressed as the probability of $k$ successes in $2E$ trials where the probability of success, i.e. creating an edge, is given by Eq.(\ref{kernel1}). $P(k|x)$ is given by a binomial variable and gives the same result when $N\rightarrow\infty$.] 

To finally reveal the fraction of nodes in the entire network of degree $k$, $p_{k}=n_{k}/N$, we need to solve the integral
\begin{equation}
\label{problem}
p_{k}=\int_{0}^{\infty}\rho(x)P(k|x)dx=\frac{l^{k}}{k!}\int_{0}^{\infty}\rho(x)x^{k}e^{-lx}dx.
\end{equation}
This is as far as a the general solution can be taken but the solutions for two special forms of $\rho(x)$ are presented in Section \ref{examples}.

\subsection{Dynamic model with memory}
The rule determining whether a node is active at any given time can be divided into two constituent mechanisms: one is regarded as a reaction to one or more previous interactions; it is memory dependent and is responsible for bursts of activity. The other is the fitness of the node which encompasses all the other reasons why a node may become active at any given time. Modifying the model (Eq.(\ref{rate1})) for the new kernel Eq.(\ref{memory_kernel}) we now have
\begin{equation}
\label{rate2}
\begin{split}
\frac{\partial n_{k}(x,t)}{\partial t}&=\\
\frac{2}{N(\langle k \rangle + \langle x \rangle)}&[(k+x-1)n_{k-1}(x,t)-(k+x)n_{k}(x,t)]\\
&+\frac{1}{E}[(k+1)n_{k+1}(x,t)-kn_{k}(x,t)].
\end{split}
\end{equation}
As before, we set the left hand side to zero to get a difference equation
\begin{equation}
\label{difference}
\begin{split}
kn_{k}&(x)-(k+1)n_{k+1}(x)=\\
&m[(k-1)n_{k-1}(x)-kn_{k}(x)+xn_{k-1}(x)-xn_{k}(x)],
\end{split}
\end{equation}
where $m=\langle k \rangle/(\langle k \rangle+\langle x \rangle)$. To solve this we introduce the generating function,
\begin{equation}
g(z,x)=\sum_{k=0}^{\infty}n_{k}(x)z^{k}
\end{equation}
by multiplying Eq.(\ref{difference}) by $z^{k}$ and summing over all $k\geq 0$ we arrive at
\begin{equation}
(z-1)(1-mz)\frac{\partial g(z,x)}{\partial z}-mx(z-1)g(z,x)=0
\end{equation}
which has the solution $g(z,x)=[C(1-mz)]^{-x}$ (a general description of this method is described in the appendix of \cite{colman2014local}). We find $C$ by substituting $g(1,x)=N(x)$ into the solution and get
\begin{equation}
\label{gen1}
g(z,x)=N(x)\left(\frac{1-m}{1-mz}\right)^{x}.
\end{equation}
The coefficient of $z^{k}$ in the expansion of the right hand side is $n_{k}(x)$, dividing this by $N(x)$ then gives the following conditional probability which contrasts with Eq.(\ref{conditional1})
\begin{equation}
\label{given}
P(k|x)=\binom{x+k-1}{k}(1-m)^{x}m^{k}.%=\frac{(1-m)^{x}m^{k}}{k!}\frac{\Gamma(x+k)}{\Gamma(x)}.
\end{equation}
As $\langle x \rangle \rightarrow \infty$, $P(k|x)$ tends towards the Poisson distribution with the same mean we had in Eq.(\ref{conditional1}). This is expected since in this limit the attachment kernel for any given node will be dominated by its fitness. Let $p_{k}$ be the fraction of nodes with degree $k$ and is the integral of the product of $\rho(x)$ and the right hand side of Eq.(\ref{given}) over all possible values of $x$
\begin{equation}
p_{k}=\frac{m^{k}}{k!}\int_{0}^{\infty}x(x+1)...(x+k-1)(1-m)^{x}\rho(x)dx.
\end{equation}
We can simplify the integral by multiplying out all the brackets which contain $x$, this gives
\begin{equation}
\label{stirling}
p_{k}=\frac{m^{k}}{k!}\sum_{n=0}^{k}c(k,n)\int_{0}^{\infty}x^{n}(1-m)^{x}\rho(x)dx.
\end{equation}
Here $c(k,n)$ denotes the unsigned Stirling numbers of the first kind (the number of permutations of $k$ symbols that have exactly $n$ cycles \cite{absteg}), since an explicit expression for these is not known, Eq.(\ref{stirling}) is only useful at small values of $k$. For large $k$ we examine the generating function
\begin{equation}
\label{series}
G(z)=\sum_{k=0}^{\infty}p_{k}z^{k}.
\end{equation}
It follows from Eq.(\ref{gen1}) that
\begin{equation}
\label{big_gen}
G(z)=\int_{0}^{\infty}\rho(x)\left(\frac{1-m}{1-mz}\right)^{x}dx.
\end{equation}
When the fitness parameter is the same for all nodes,  $x_{i}=\alpha$, the model reduces to that studied in \cite{xie2008scale}. Substituting $\rho(x)=\delta(x-\alpha)$ into Eq.(\ref{big_gen}) yields the expected result.

\section{Examples of specific fitness distributions}
\label{examples}
\subsection{Gamma distribution}
\label{gamma_section}
We examine in detail the possible scenario where the fitness of the population follows the gamma distribution 
\begin{equation}
\label{gamma_dist}
\rho(x;\alpha,\beta)=\frac{x^{\alpha-1}e^{-x/\beta}}{\beta^{\alpha}\Gamma(\alpha)}
\end{equation}
which generalises a number of distributions that have applications in social sciences including $\chi^{2}$ and the exponential distribution. In general it has the appearance of an asymmetric bell curve and we consider it entirely likely that a system might exist where the fitness values are clustered around the mean in this way. 

\subsubsection{Dynamic model without memory} 
We solve Eq.(\ref{problem}) to find the degree distribution. The integral becomes
\begin{equation}
p_{k}=\frac{l^{k}}{k!\beta^{\alpha}\Gamma(\alpha)}\int_{0}^{\infty}x^{k+\alpha-1}e^{-x(l+1/\beta)}dx.
\end{equation}
By applying the change of variables $y=(l+1/\beta)x$ the integral becomes the product of a gamma function and some other factors. We arrive at
\begin{equation}
p_{k}=\frac{l^{k}\Gamma(k+\alpha)}{\beta^{\alpha}(l+1/\beta)^{k+\alpha}\Gamma(\alpha)k!}.
\end{equation}
For large values of $k$ this solution becomes a gamma distribution $p_{k}\sim\rho(k;\alpha,\log(1+1/l\beta))$.
 
\subsubsection{Dynamic model with memory} 
We substitute Eq.(\ref{gamma_dist}) into Eq.(\ref{stirling}) and applying the change of variables; $y=x[(1/\beta)-\log(1-m)]$, we can again take a gamma function out as a factor, leaving
\begin{equation}
p_{k}=\frac{m^{k}}{k!\beta^{\alpha}\Gamma(\alpha)}\sum_{n=0}^{k}c(k,n)\Gamma(n+\alpha)\left[\log\left(\frac{1}{1-m}\right)+\frac{1}{\beta}\right]^{-(n+\alpha)}.
\end{equation}
Substituting Eq.(\ref{gamma_dist}) into Eq.(\ref{big_gen}) and solving the integral we arrive at
\begin{equation}
G(z)=\left[1+\log\left(\frac{1-mz}{1-m}\right)^{\beta}\right]^{-\alpha}.
\end{equation}
As $z\rightarrow 1$, the logarithm in the above expression approaches $0$ making the approximation $\log(X)\approx X-1$ appropriate to use. For $z\approx 1$ we have
\begin{equation}
G(z)\approx\left(\frac{1-m}{1-mz}\right)^{\alpha\beta}
\end{equation}
which can be expanded to recover the power series. Equating the coefficients of the expansion with those of Eq.(\ref{series}) we find
\begin{equation}
p_{k}\approx(1-m)^{\alpha\beta}\binom{-\alpha\beta}{k}(-m)^{k}
\end{equation}
For large $k$ this is
\begin{equation}
\label{gamma_result}
p_{k}\approx\frac{(1-m)^{\alpha\beta}}{\Gamma(\alpha\beta)}m^{k}k^{\alpha\beta-1}=c\rho\left(k;\alpha\beta,\frac{1}{\log(1/m)}\right)
\end{equation}
where $c=\left[(1-m)/\log(1/m)\right]^{\alpha\beta}$ is a normalising constant. As the mean $\langle x \rangle=\alpha\beta$ tends towards $0$ the distribution tends towards a power-law with exponent $-1$. This represents a scenario where the majority of actions are in fact reactions to previous events. 

\subsection{Power-law distribution}
\label{power_section}
Suppose fitness is distributed according to the following power-law
\begin{equation}
\label{power_dist}
 \rho(x; x_{\text{min}},\gamma) =\left\{
   \begin{aligned}
      \frac{\gamma-1}{x_{\text{min}}}&\left(\frac{x}{x_{\text{min}}}\right)^{-\gamma} & \quad \text{if}\quad x\geq x_{\text{min}}\\
     &0 & \quad \text{if}\quad x<x_{\text{min}}
   \end{aligned}\right.
\end{equation}
which has the mean
\begin{equation}
\label{power_mean}
\langle x \rangle=\frac{\gamma-1}{\gamma-2}x_{\text{min}}.
\end{equation}

\subsubsection{Dynamic model without memory} 
To find the degree distribution we substitute Eq.(\ref{power_dist}) into  Eq.(\ref{problem}), giving
\begin{equation}
\label{power_integral}
p_{k}=\frac{(\gamma-1)l^{k}}{x_{\text{min}}^{1-\gamma}k!}\int_{x_{\text{min}}}^{\infty}x^{k-\gamma}e^{-lx}dx.
\end{equation}
Using the substitution $y=lx$, the integral can be expressed using the upper incomplete gamma function (see \cite{absteg}) defined as $\Gamma(u,v)=\int_{v}^{\infty}\upsilon^{u-1}e^{-\upsilon}d\upsilon$ for real numbers $u$ and $v$. We can also simplify the solution by combining the parameters using
\begin{equation}
\label{A}
A=lx_{\text{min}}=\frac{2E(\gamma-2)}{N(\gamma-1)}
\end{equation}
and we get
\begin{equation}
\label{power_no_memory}
p_{k}=\frac{(\gamma-1)A^{\gamma-1}}{k!}\Gamma(k-\gamma+1,A).
\end{equation}
Notice that all choices of $x_{\text{min}}$ yield the same result. This is not unexpected; the scale-invariance of the power law distribution means that generating a random fitness $x_{i}$ using Eq.(\ref{power_dist}) is eqivalent to generating $\xi$ from $\rho(\xi,1,\gamma)$ and taking $x_{i}=\xi x_{\text{min}}$ as the fitness value. Substituting this fitness value into Eq.(\ref{kernel1}) we see that $x_{\text{min}}$ is no longer present. 

It is also informative to solve Eq.(\ref{power_integral}) for integer values of $\gamma$. We first express the part inside the integral as a derivative
\begin{equation}
x^{k-\gamma}e^{-lx}=\left.(-1)^{k-\gamma}\frac{d^{k-\gamma}e^{-xy}}{dy^{k-\gamma}}\right|_{y=l}
\end{equation}
before performing the integration with respect to $x$. Since
\begin{equation}
\int_{x_{\text{min}}}^{\infty}e^{-xy}dx=\frac{\exp(-x_{\text{min}}y)}{y}
\end{equation}
and 
\begin{equation}
\begin{split}
\frac{d^{n}}{dy^{n}}&\left(\frac{\exp(-x_{\text{min}}y)}{y}\right)\\
&=(-1)^{n}\exp(-x_{\text{min}}y)\sum_{s=0}^{n}\frac{n!}{s!}x_{\text{min}}^{s}y^{s-n-1}
\end{split}
\end{equation}
for $n \in \mathbb{N}$, for integer values of $\gamma$ we arrive at
\begin{equation}
p_{k}=\frac{(\gamma-1)(lx_{\text{min}})^{\gamma-1}\exp(-lx_{\text{min}})}{k(k-1)\ldots(k-\gamma+1)}\sum_{s=0}^{k-\gamma}\frac{(lx_{\text{min}})^{s}}{s!},
\end{equation}
which, using Eq.(\ref{A}) simplifies to
\begin{equation}
\label{power_sol1}
p_{k}=\frac{(\gamma-1)A^{\gamma-1}e^{-A}}{k(k-1)\ldots(k-\gamma+1)}\sum_{s=0}^{k-\gamma}\frac{A^{s}}{s!}.
\end{equation}
It is now easy to see that the degree distribution has a power-law tail (see Fig.(\ref{power_forgetful})).

\subsubsection{Dynamic model with memory}
First we substitute Eq.(\ref{power_dist}) into Eq.(\ref{stirling}). We introduce $L=-\log(1-m)$, then, by applying a the change of variables $y=Lx$, we can factorise out an incomplete gamma function. This gives the following exact solution for the degree distribution
\begin{equation}
\label{power_law_small}
p_{k}=\frac{m^{k}(\gamma-1)}{k!x_{\text{min}}^{1-\gamma}}\sum_{n=0}^{k}c(k,n)L^{\gamma-n-1}\Gamma(n-\gamma+1,Lx_{\text{min}}).
\end{equation}
The parameter $x_{\text{min}}$, which was absent in Eq.(\ref{power_no_memory}), now controls the overall effect of fitness in proportion to memory. For large values of $k$ we solve Eq.(\ref{big_gen}) to find 
\begin{equation}
G(z)=(\gamma-1)\Gamma[1-\gamma,\Phi(x_{\text{min}},z)][\Phi(x_{\text{min}},z)]^{\gamma-1}
\end{equation}
where 
\begin{equation}
\Phi(x_{\text{min}},z)=x_{\text{min}}\log\left(\frac{1-mz}{1-m}\right).
\end{equation}
Using the approximation $\log(X)\approx X-1$ as $z$ approaches $1$ we find
\begin{equation}
G(z)\approx(\gamma-1)^{\gamma-1}\Gamma(1-\gamma)A^{\gamma-1}(1-z)^{\gamma-1}
\end{equation}
where $A$ is given by Eq.(\ref{A}). For non-integer values of $\gamma$ this can be expanded and the coefficients of the expansion can be equated with Eq.(\ref{series}). We see that
\begin{equation}
p_{k}\approx\frac{(\gamma-1)A^{\gamma-1}\Gamma(k-\gamma+1)}{\Gamma(k+1)}.
\end{equation}
The $k$ dependence exists in the form of the ratio of two gamma functions so asymptotically $p_{k}\sim k^{-\gamma}$. It is worth remarking that the power-law exponent in the fitness distribution is the same exponent found in the degree distribution and is not affected by the choice of the other parameters $N$, $E$ or $x_{\text{min}}$ as can be seen in Fig.(\ref{power_memory}).

\subsection{Step function distribution}
For practical purposes it is useful to have a general method of inferring a fitness distribution from a degree distribution. We suggest one such approach here and focus exclusively on the case where memory effects are present.

 By assuming the fitness distribution has the form of a step function (otherwise know as a staircase function) we can minimise the error between the theoretical prediction and the observed data by adjusting the height of each step (or stair). Suppose we have a vector of parameters $\textbf{a}=[a_{0}, a_{1},...,a_{J}]$, we then define the distribution as
\begin{equation}
\rho(x,\textbf{a},\delta)=\left\{ 
\begin{aligned} 
    a_{0} & \quad \text{for } 0<x\leq\delta \\
    a_{1} & \quad \text{for } \delta<x\leq2\delta\\
    \vdots & \\
    a_{j} & \quad \text{for } j\delta<x\leq(j+1)\delta\\
    \vdots & \\
    a_{J} & \quad \text{for } J\delta<x\leq(J+1)\delta
  \end{aligned} \right.
\end{equation}
where $\sum_{i=0}^{J}a_{j}=[\delta J]^{-1}$. The mean fitness is 
\begin{equation}
\langle x \rangle=\textbf{a}\textbf{v}
\end{equation}
where $\textbf{v}=(\delta/2)[1,3,...,2J+1]$. Substituting $\rho$ into Eq.(\ref{big_gen}) we get
\begin{equation}
\label{step1}
\begin{split}
G(z)=&\sum_{i=0}^{I}a_{i}\int_{i\delta}^{(i+1)\delta}\left(\frac{1-m}{1-mz}\right)^{x}dx\\
=&\frac{\left[(1-m)/(1-mz)\right]^{\delta}-1}{\log[(1-m)/(1-mz)]}\sum_{i=0}^{I}a_{i}\left(\frac{1-m}{1-mz}\right)^{\delta i}.
\end{split}
\end{equation}
We can generate (randomly or systematically) a vector of values $\textbf{z}=[z_{0}, z_{1},...,z_{I}]$ at which the generating function can be evaluated. The empirical data is the degree distribution $\textbf{p}=[p_{0},p_{1},...,p_{K}]$ where $K$ is the largest degree. The degree distribution $\textbf{p}$, the generating function as given by Eq.(\ref{series}), fitness parameters $\textbf{a}$, and the generating function as given by Eq.(\ref{step1}) are all connected by the following expression:
\begin{equation}
\label{matrix_eq}
Z\textbf{p}=W\textbf{a}.
\end{equation}
Here $Z$ is a $I\times K$ matrix whose $(i,k)$th entry is $z_{i,k}=z_{i-1}^{k-1}$ and $W$ is a $I\times J$ matrix whose $(i,j)$th entry is given by
\begin{equation}
w_{i,j}=\frac{[(1-m)/(1-mz_{i-1})]^{\delta}-1}{\log[(1-m)/(1-mz_{i-1})]}\left(\frac{1-m}{1-mz_{i-1}}\right)^{\delta (j-1)}.
\end{equation}
While Eq.(\ref{matrix_eq}) appears to be a simple linear algebra problem, it is complicated by the fact that $m$ depends on $\langle x\rangle$, which is only known \emph{after} a choice of $\textbf{a}$ has been made, therefore $W$ is a function of both $\textbf{a}$ and $\delta$. This does however provide a neat way to formally present the problem: We choose $J<K$ to prevent having more parameters than datapoints and solve
\begin{equation}
\label{opt}
\rho(x)=\rho(x,\textbf{a},\delta)
\end{equation}
such that 
\begin{equation}
 \left\|Z\textbf{p}-W\textbf{a}\right\| = \min_{\textbf{a},\delta}\left\|Z\textbf{p}-W\textbf{a}\right\|.
\end{equation}
%\vspace{3cm}
\newpage

\bibliography{bibfile2}

\end{document}